\documentclass[twocolumn,prb,showpacs,amsmath,amssymb]{revtex4}
\usepackage{graphicx}
\usepackage{dcolumn}
\usepackage{bm}
\begin{document}

\title {Magnetic proximity effect in the 3D topological insulator/ferromagnetic insulator heterostructure}
\author{V. N. Men'shov$^{1,2,3}$}
\author{V. V. Tugushev$^{1,2,3,4}$}
\author{S. V. Eremeev$^{1,3,5}$}
\author{P. M. Echenique$^{1,6}$}
\author{E. V. Chulkov$^{1,6}$}
\affiliation{$^1$Donostia International Physics Center (DIPC), P.
de Manuel
Lardizabal 4, 20018, San Sebasti{\'a}n, Basque Country, Spain\\
$^2$NRC Kurchatov Institute, Kurchatov Sqr. 1, 123182 Moscow, Russia\\
$^3$Tomsk State University, 634050 Tomsk, Russia\\
$^4$Prokhorov General Physics Institute, Vavilov Str. 38, 119991, Moscow, Russia\\
$^5$Institute of Strength Physics and Materials Science, 634021,
Tomsk, Russia\\
$^6$Departamento de F\'{\i}sica de Materiales UPV/EHU, CFM - MPC
UPV/EHU,
20080 San Sebasti\'an/Donostia, Basque Country, Spain\\
}

\date{\today}

\begin{abstract}
We theoretically study the magnetic proximity effect in the three
dimensional (3D) topological insulator/ferromagnetic insulator
(TI/FMI) structures in the context of possibility to manage the
Dirac helical state in TI. Within continual approach based on the
$\mathbf{kp}$ Hamiltonian we predict that, when 3D TI is brought
into contact with 3D FMI, the ordinary bound state arising at the
TI/FMI interface becomes spin polarized due to the orbital mixing
at the boundary. Whereas the wave function of FMI decays into the
TI bulk on the atomic scale, the induced exchange field, which is
proportional to the FMI magnetization, builds up at the scale of
the penetration depth of the ordinary interface state. Such the
exchange field opens the gap at the Dirac point in the energy
spectrum of the topological bound state existing on the TI side of
the interface. We estimate the dependence of the gap size on the
material parameters of the TI/FMI contact.
\end{abstract}


\maketitle

\section{INTRODUCTION}

The engineering of layered hybrid heterostructures for spintronics
applications, which are apt to simultaneously generate, manipulate
and detect electron spin-polarized currents, sends a formidable
challenge to modern material science and technology.\cite{Wolf}
The information obtained from combining traditional semiconductors
with magnetic materials indicates the key influence of the
interface between the constituents on magnetic properties of the
heterostructure.\cite{Zutic,Jansen} The electronic rearrangement
at the interface, often accompanied by appearance of
spin-polarized bound states on both sides of the boundary, is the
central feature inherent in the hybrid heterostructures. This
feature underlies the magnetic proximity effect when spin ordering
penetrates inside semiconductor over the interface.\cite{Men-10}
Naturally one asks what are the peculiarities of the magnetic
proximity effect provided that a trivial band semiconductor in the
hybrid heterostructure is replaced by a time-reversal-invariant
semiconductor with inverted band gap driven by strong intrinsic
spin-orbit coupling (SOC), which is now known as a three
dimensional topological insulator (3D TI).\cite{Hasan,Qi} A
characteristic feature of the heterostructures containing the 3D
TI layers is the topologically protected electron helical states
with the linear Dirac-cone-like energy-momentum dispersion living
at a boundary between topologically nonequivalent regions (such as
between TI and a conventional, i.e. topologically trivial,
insulator).\cite{Hasan,Qi} When TI is in contact with a magnetic
material, the interface states can be influenced by a presence of
magnetic ordering via time reversal symmetry breaking. Because of
the Dirac spectrum and spin-momentum locking in such the systems,
an interplay between the interface states and magnetism is
naturally expected to display novel behavior that is absent in the
case of the heterostructure composed of a conventional
semiconductor and a magnetic material. In integrating of TIs with
magnetic materials, fundamental issue is characterization of
electron properties of the interface to understand the mechanisms
of an exchange coupling.

The design of the heterostructures made of TI and magnetic layers
has the potential to achieve a strong and uniform exchange
coupling between the layers without significant spin dependent
random scattering of helical carriers on magnetic atoms. The main
challenge is to find suitable magnetic material, which can form a
high-quality contact with TI and, at the same time, provides a
strong exchange coupling at the interface. Unfortunately magnetic
metals are out of the range of the search since they naturally
short circuit the TI layer, restricting fundamentally the device
design. The film of a ferromagnetic insulator or semiconductor
(below we will use the abbreviation FMI) adjacent to TI is a most
promising candidate to manipulate the helical states of 3D TI by
means of a magnetic proximity effect.\cite{Garate} Such a way may
diminish the surface scattering via continuum states of the
magnetic film, contrary to the metallic film case. There are
several experimental investigations directly addressable to the
hybrid TI/FMI structures. Zhou \textit{et al} had studied
semiconductor trilayer structures with ferromagnetic
Sb$_{2-x}$Cr$_{x}$Te$_{3}$ layers separated by the
Sb$_{2}$Te$_{3}$ layer.\cite{Zhou} Ferromagnetic out-of-plane
exchange coupling between the Sb$_{2-x}$Cr$_{x}$Te$_{3}$ layers
was found for a sample with the Sb$_{2}$Te$_{3}$ spacer thickness
of 2 nm. Recently  Kandala \emph{et al.} studied hybrid
heterostructure composed of TI (Bi$_2$Se$_3$) and FMI (GdN) layers
to probe the effects of broken time reversal symmetry on
electrical transport in the surface states of 3D TI.\cite{Kandala}
Low temperature longitudinal magnetoconductance data presented in
Ref.~\onlinecite{Kandala} are consistent with the opening of a
magnetic gap in the surface state spectrum at the Bi$_2$Se$_3$/GdN
interface. In Ref.~\onlinecite{Ji} it is demonstrated that the
layered room temperature ferromagnet Fe$_7$Se$_8$ grows very well
between layers of Bi$_2$Se$_3$ in bulk crystals. Both phases in
the intergrown composite Bi$_2$Se$_3$:Fe$_7$Se$_8$ crystals
display their intrinsic bulk properties: the ferromagnetism of
Fe$_7$Se$_8$ is anisotropic, with magnetization easy axis in the
plane of the crystal, and ARPES characterization shows that the
topological states remain present on the Bi$_2$Se$_3$ surface. The
heterostructures comprised of the layers of hexagonal TI
Bi$_2$Se$_3$ and cubic FMI EuS with the sharp interface were
fabricated by molecular beam epitaxy in Ref.~\onlinecite{Wei}. The
magnetic and magnetotransport measurements on the Bi$_2$Se$_3$/EuS
bilayers indicated eloquently that EuS induces ferromagnetic order
with a significant magnetic moment in the interfacial region of
the Bi$_2$Se$_3$ layer due to a transmission of the exchange field
across the interface.\cite{Wei}

Of magnetic insulators that show relatively good lattice matching
with binary TIs Bi$_2$Te$_3$, Bi$_2$Se$_3$ and Sb$_2$Te$_3$, one
can note the wide-gap antiferromagnetic insulator (AFMI)
MnSe.\cite{Luo} The authors of Ref.~\onlinecite{Luo} performed
first-principles calculation for the Bi$_2$Se$_3$/MnSe
superlattice electron structure; in particular, they predicted
that exchange coupling with MnSe induces a gap of 54 meV in the
surface states spectrum of Bi$_2$Se$_3$.

There are theoretical studies of the properties of TI in contact
with a conventional FMI, which are routinely based on the simple
phenomenological Hamiltonian of the 2D Dirac-like states of
helical fermions in an homogeneous exchange field:\cite{Fu}
$H_{s}=-iv(\mathbf{e}_z[\bm{\sigma}\times
\bm{\bigtriangledown}])+\mathbb{U}_{ex}$, where $\mathbf{e}_z$ is
the unit vector normal to the interface, $v$ is the Fermi
velocity. It is generally thought that a controllable exchange
field on the TI side of the TI/FMI hybrid structure can be
directly generated by means of a FMI layer attached to a TI
layer.\cite{Garate,Tserkovnyak,Yokoyama} In such the case one
presumes $\mathbb{U}_{ex}=J(\bm{\sigma}\mathbf{M})$, i.e., the
exchange field is proportional to the FMI magnetization
$\mathbf{M}$ on the FMI side of the structure, $J$ is effective
exchange coupling and $\bm{\sigma}$ is the vector composed of the
Pauli matrices. Within this description, it was predicted that
many curious effects can be realized in the TI/FMI structures.
However, strictly speaking, the 2D Hamiltonian $H_{s}$ can
formally be derived from a relevant 3D Hamiltonian only under the
stipulation that TI has a free surface on which
$\mathbb{U}_{ex}=0$. Usually, to take into account a perturbation
from the interface with magnetic material, the exchange term is
simply included in the Hamiltonian $H_{s}$, without a serious
analysis of its microscopic origin. The same one can refer to an
attempt to go beyond the scope of the 2D model.\cite{Habe}

In the present paper, within the framework of the continual
approach, we study the physics of magnetic proximity effect at the
TI/FMI heterocontact. This model generalizes an approach recently
proposed to describe the electron states formed by the interface
between TI and normal insulator (NI).\cite{Men-13} As was shown in
Ref.~\onlinecite{Men-13} the boundary between TI and a trivial
insulator hosts a pair of the bound electron states inside the
bulk energy gap of TI on the topological side of the interface
(referred as the topological and ordinary states), which differ
from each other in physical meaning, spatial distribution and
energy spectrum. Namely, the bound topological state stems from a
breaking of the $Z_2$ topological invariant at the boundary with a
trivial insulator. This state being located relatively remote from
the interface and almost insensitive to the interface influence
shows the Dirac spectrum. By contrast, the bound ordinary state
results from the crystal symmetry breaking at the interface. This
state is spatially located near the interface, therefore, its
features strongly depend on the effective interface potential.
Here we generalize the approach of Ref.~\onlinecite{Men-13} to the
case when TI is attached to FMI. Our analysis sheds light on the
origin of the proximity effect in the TI/FMI structures and the
possibility of magnetic control over the Dirac helical state in
TI.

The paper is organized as follows. In Sec. II, we discuss the
model for a contact between TI and FMI and introduce the main
ingredients and assumptions of the problem within the continual
approach, which takes into account the hybridization between the
orbitals of the constituents at the interface. In Sec. III, we
argue the occurrence of the spin-polarized ordinary interface
state on the topological side of the contact, derive the spin
polarization of carriers in the ordinary state that is induced by
the adjacent FMI and determine the dependence of the polarization
on the distance from the interface. In Sec. IV the behavior of the
topological interface state under the influence of the exchange
field associated with the ordinary state is described. We show how
the proximity of the FMI modifies the energy spectrum of the
topological state and obtain the expression for the energy gap. In
Sec. V, we analyze the obtained results and compare them with the
recent \textit{ab-initio} calculations for the Bi$_2$Se$_3$/MnSe
superlattice.\cite{Eremeev}

\section{MODEL HAMILTONIAN}

We consider a conceptual analytic model for the magnetic proximity
effect in the TI/FMI structure. The low energy and long wavelength
bulk electron states of the prototypical TI, narrow-gap
semiconductor of Bi$_2$Se$_3$-type, are described near the
$\Gamma$ point of the Brillouin zone by the four bands
$\mathbf{kp}$ Hamiltonian with strong SOC proposed in
Ref.~\onlinecite{Liu}. Without a loss of generality, we make use
the simplified version of this Hamiltonian in the form:
\begin{equation}\label{H-TI}
H_{t}^{0}(\mathbf{k})=\Xi(\mathbf{k})\tau_{z}\otimes\sigma_{0}
+\mathrm{A}(k_{x}\tau_{x}+k_{y}\tau_{y})\otimes\sigma_{x}+
\mathrm{A}k_{z}\tau_{x}\otimes\sigma_{z},
\end{equation}
where $\Xi(\mathbf{k})=\Xi-\mathrm{B}k^{2}$, $\mathbf{k}$ is the
wave vector, $k=|\mathbf{k}|$, $\sigma_{\alpha}$ and
$\tau_{\alpha}$ ($\alpha=0,x,y,z$) denote the Pauli matrices in
the spin and orbital space, respectively. The condition
$\Xi,\mathrm{B}>0$ reflects an inverted order of the energy terms
around the $\Gamma$ point $\mathbf{k}=0$ as compared with large
$k$, which correctly characterizes the topologically non-trivial
nature of the system due to strong SOC.

For the sake of simplicity we introduce the FMI as a wide-gap
semiconductor in which the exchange potential of local magnetic
moments induces uniform spin splitting of both the conduction and
valence bands in the bulk. Thus, the low energy and long
wavelength bulk electron states of FMI are formally modeled by the
four bands $\mathbf{kp}$ Hamiltonian without SOC:
\begin{equation}\label{H-MI}
H_{f}^{0}(\mathbf{k})=\mathrm{E}_{0}(\mathbf{k})\mathbb{I}+
\Lambda(\mathbf{k})\tau_{z}\otimes\sigma_{0}+\mathrm{M}\tau_{0}\otimes\sigma_{z},
\end{equation}
where $\mathbb{I}$ is unit $4\times4$ matrix. For the sake of
convenience we use a simple effective mass approximation so that
$\Lambda(\mathbf{k})=\Lambda+\mathrm{N}k^{2}$,
$\Lambda,\mathrm{N}>0$. The band structure (\ref{H-MI}) is
generally asymmetric with respect to the middle of the TI band
gap, since $\mathrm{E}_{0}(\mathbf{k})\neq0$; in the following we
omit for simplicity the $\mathbf{k}$ dependence of
$\mathrm{E}_{0}(\mathbf{k})$ and put $\mathrm{E}_{0}(\mathbf{k}) =
\mathrm{E}_{0}$. The intrinsic magnetization of FMI is assumed
perpendicular to the TI/FMI interface plane. We regard FMI as a
wide-gap semiconductor in the sense that
$|\mathrm{E}_{c,v}^{\sigma}|>\Xi$, where
$\mathrm{E}_{c}^{\sigma}=\mathrm{E}_{0}+\Lambda+\sigma\mathrm{M}$
and $\mathrm{E}_ {v}^ {\sigma}=
\mathrm{E}_{0}-\Lambda+\sigma\mathrm{M}$ are the edges of the
conduction and valence bands, respectively,
$\Lambda,\mathrm{M}>0$, $\sigma=\pm1$ is spin projection on the
quantization axis $z$.

We assume that the TI occupies the right half-space, $z>0$, while
the FMI occupies the left one, $z<0$. Both the TI and FMI are
treated as semi-infinite materials joined at a perfectly flat
interface located at $z=0$. In a real space coordinate
representation, the electron energy of the TI/FMI contact reads:
\begin{eqnarray}\label{Omega0}
\Omega&=&\int_{z>0} d\mathbf{r} \Theta^{\dag}(\mathbf{r})H_{t}
(-i\nabla)\Theta(\mathbf{r})\\&+& \int_{z<0} d\mathbf{r}
\Phi^{\dag}(\mathbf{r})H_{m} (-i\nabla)\Phi(\mathbf{r})+\Omega_{I}
\nonumber,
\end{eqnarray}
\begin{equation}\label{H-T-til}
H_{t}(-i\nabla)=H_{t}^{0}(-i\nabla)+\varphi(\mathbf{r})\mathbb{I}+
\tau_{0}\otimes(\bm{\sigma}\cdot\bm{\Delta}(\mathbf{r})),
\end{equation}
\begin{equation}\label{H-M-til}
H_{f}(-i\nabla)=H_{f}^{0}(-i\nabla)+\varphi_{m}(\mathbf{r})\mathbb{I}+
\tau_{0}\otimes(\bm{\sigma}\cdot\bm{\Delta}_{m}(\mathbf{r})),
\end{equation}
\begin{equation}\label{Omega-I}
\Omega_{I}=\int
d\mathbf{r}[\Theta^{\dag}(\mathbf{r})V(\mathbf{r})\Phi(\mathbf{r})+
\Phi^{\dag}(\mathbf{r})V^{\dag}(\mathbf{r})\Theta(\mathbf{r})],
\end{equation}

Here the operators $H_{t}^{0}(-i\nabla)$ and $H_{f}^{0}(-i\nabla)$
determined in Eqs. (\ref{H-TI})-(\ref{H-MI}) (momentum
$\mathbf{k}$ is replaced by operator $-i\nabla$) act in the space
of the spinor envelope functions
$\Theta(\mathbf{r})=(\theta_{1}(\mathbf{r}),
\theta_{2}(\mathbf{r}),\theta_{3} (\mathbf{r}), \theta_{4}
(\mathbf{r}))^{\mathrm{T}}$ and
$\Phi(\mathbf{r})=(\phi^{(1)}(\mathbf{r}),\phi^{(2)}(\mathbf{r}),
\phi^{(3)}(\mathbf{r}),\phi^{(4)}(\mathbf{r}))^{\mathrm{T}}$,
respectively. The subscript $j$ indicates the TI state spinor
components, $\theta_{j}(\mathbf{r})$, while the superscript $(n)$
indicates the FMI state spinor components,
$\phi^{(n)}(\mathbf{r})$.

Since TI is a narrow-gap semiconductor, it is evident that a
significant electric field can be induced on the TI side of the
TI/FMI contact due to the redistribution of the charge density of
carriers $n(\mathbf{r})$ in the sub-interface layers of TI. This
field contains, in principle, the components of different spatial
scales. In our approach, the short-range (of the order of
inter-atomic distance) components of the Coulomb potential are
assumed to be included in the spin-independent terms of an
effective local interface potential (see below). The long-range
profile of the Coulomb potential $\varphi(\mathbf{r})$ in
Eq.~(\ref{H-T-til}) near the interface can be formally described
by the equation
\begin{equation}\label{phi}
\varphi(\mathbf{r})=\int
d\mathbf{r}'\mathcal{V}(\mathbf{r}-\mathbf{r}') n(\mathbf{r}'),
\end{equation}
where $\mathcal{V}(\mathbf{r})$ is the spin-independent part of
electron-electron interaction in TI. Besides, a noticeable
redistribution of the carrier spin density $s(\mathbf{r})$ can
appear in TI due to exchange coupling between the states of TI and
FMI at the interface. This redistribution induces an exchange
field in the sub-interface layers of TI. In our approach, the
short-range components of this field are included into the
spin-dependent  terms of an effective local interface potential
(see below), while the relative long-range components of the spin
density cause the spin-dependent field $\bm{\Delta}(\mathbf{r})$
in Eq.~(\ref{H-T-til}):
\begin{equation}\label{del}
\Delta^{\alpha}(\mathbf{r})=\int
d\mathbf{r}'\mathcal{K}^{\alpha\beta}(\mathbf{r}-\mathbf{r}')
s^{\beta}(\mathbf{r}').
\end{equation}
The diagonal elements of the matrix
$\mathcal{K}^{\alpha\beta}(\mathbf{r})$ with $\alpha=\beta$ are
due to the exchange term of the electron-electron interaction in
TI, while the off-diagonal elements with $\alpha\neq\beta$ appear
due to the spin-flip electron-electron scattering and vanish
without SOC.

In general case, not only the bound interface states under the
study, but all engaged electron states of TI give rise to the
charge and spin densities, $n(\mathbf{r})$ and $s(\mathbf{r})$.
The fields $\varphi(\mathbf{r})$ and $\bm{\Delta}(\mathbf{r})$ in
Eq.~(\ref{H-T-til}) are associated with the band bending and spin
splitting on the TI side of the contact, respectively. It is clear
that similar expressions can be formally written for the
potentials $\varphi_{f}(\mathbf{r})$ and $\bm{\Delta}_{f}
(\mathbf{r})$ on the MI side of the contact. However, since FMI is
a wide-gap semiconductor, it is reasonable to assume in
Eq.~(\ref{H-M-til}) that $|\varphi_{f} (\mathbf{r})|<<2\Lambda$
and $|\bm{\Delta}_{f} (\mathbf{r})|<<2\mathrm{M}$, thus neglecting
the effect of charge and spin redistributions for the FMI
subsystem.

To formally make the potentials $\varphi(\mathbf{r})$ and
$\Delta^{\alpha}(\mathbf{r})$ [as well as the interactions
$\mathcal{V}(\mathbf{r})$ and
$\mathcal{K}^{\alpha\beta}(\mathbf{r})$] consistent with the
densities $n(\mathbf{r})$ and $s(\mathbf{r})$ it is necessary to
solve the system of the Dyson equations for the self-energy parts
of the Green`s functions on the TI side of the contact. One can
hardly accomplish this task analytically, so we make some
approximations. First, the matrices $\mathcal{V}(\mathbf{r})$ and
$\mathcal{K}^{\alpha\beta} (\mathbf{r})$ are supposed to be
independent of the orbital indices. Second, we use the "local"
approximation for an electron-electron interaction:
$\mathcal{V}(\mathbf{r})\rightarrow \mathcal{V}\delta(\mathbf{r})$
and $\mathcal{K}^{\alpha\beta}(\mathbf{r})
\rightarrow\mathcal{K}^{\alpha\beta}\delta(\mathbf{r})$
[$\delta(\mathbf{r})$ is the delta-function], which is specific to
a metal situation. For the Coulomb interaction such the assumption
is not evident, since the effective scale of
$\mathcal{V}(\mathbf{r})$ defined by the Debye screening length,
$D$, may significantly exceed a characteristic metal screening
length due to a relatively low carrier concentration in TI.
Nevertheless, we formally suggest that, near the interface, the
concentration $n(\mathbf{r})$ is large enough to efficiently
screen the interaction between carriers. As for the exchange and
spin-orbit components, the scale of
$\mathcal{K}^{\alpha\beta}(\mathbf{r})$ is at least one order of
magnitude lower as compared to $D$, hence in this case the "local"
approximation is correct. Third, we average the redistributions
$n(\mathbf{r})$ and $s(\mathbf{r})$ over the $(x,y)$ plane
remaining one-dimensional profiles $n(z)$ and $s(z)$. As a result
of the aforesaid approximations we arrive at the following
relations $\varphi(z)=\mathcal{V}n(z)$ and
$\bm{\Delta}^{\alpha}(z)=\mathcal{K}^ {\alpha\beta}s^ {\beta}(z)$.
This means that, in Eq.~(\ref{H-T-til}), we consider electrons
under the one-dimensional potential fields, $\varphi(z)$ and
$\bm{\Delta}^{\alpha}(z)$, smoothly varying (on an atomic scale)
in the $z$ direction and homogeneous along the $(x,y)$ plane.

The $\mathbf{kp}$ method cannot provide information on the
wave-function behavior in the vicinity of the atomically sharp
interface, where large momenta are highly important. To overcome
this drawback we bring in the effective potential of hybridization
$V(\mathbf{r})$, which intermixes the TI and FMI electron states
at the interface. The hybridization potential $V(\mathbf{r})$
spreads over a small region $d$ (of the order a lattice parameter)
around the geometrical boundary $z=0$, where the $\mathbf{kp}$
scheme is not valid. An introduction of the phenomenological term
of the interface energy $\Omega_{I}$ enables us to correctly
reconcile the short-range (at $|z|<d<<D$) and long-range (at
$|z|>d$) variations of the charge and spin densities near the
interface in terms of the boundary conditions for the envelope
functions $\Theta(\mathbf{r})$ and $\Phi(\mathbf{r})$. The
influence of the short-range variations of $n(\mathbf{r})$ and
$s(\mathbf{r})$ is implied to be included into the effective
potential of the hybridization. As long as the spatial variations
of the interface states are sufficiently slow on the scale of the
length $d$, one can adopt for practical calculations a local
approximation for the hybridization potential,
$V(\mathbf{r})=dV(x,y)\delta(z)$.

\section{SPIN POLARIZED ORDINARY BOUND STATE}

Since the system under consideration displays translational
symmetry in the interface plane, it is reasonable to use the mixed
$(\bm{\kappa},z)$ representation, where $\bm{\kappa}$ is 2D wave
vector in the $(x,y)$ plane. We are interested in obtaining the
eigen states of the problem, Eqs.~(\ref{Omega0})-(\ref{Omega-I}),
with decaying asymptotics far from the interface,
$\Phi(\bm{\kappa},z\rightarrow -\infty)$ and
$\Theta(\bm{\kappa},z\rightarrow \infty)$. Recently, within the
variational approach, it was shown\cite{Men-13} that the energy
functional Eq.~(\ref{Omega0}) possesses two different extremals on
the class of piecewise smooth in both half-spaces and square
integrable functions. In other words, for each $\kappa$-mode, one
can write the spinor envelope functions
$\{\Phi(\bm{\kappa},z),\Theta(\bm{\kappa},z)\}$ satisfying the
same Euler equation,
$[H_{m}(\bm{\kappa},-i\partial_{z})-E]\Phi(\bm{\kappa},z)=0$ at
$z<0$ and
$[H_{t}(\bm{\kappa},-i\partial_{z})-E]\Theta(\bm{\kappa},z)=0$ at
$z>0$ (where $\partial_{z}=\partial/\partial z$), but distinct
boundary conditions at the interface $z=0$. On the one hand, one
can strictly fix the magnitude of the envelope functions at the
interface, $\{\Phi(\bm{\kappa},0-)=0,\Theta(\bm{\kappa},0+)=0\}$
to obtain the so-called interface topological state
$\{\Phi_{t}(\bm{\kappa},z),\Theta_{t}(\bm{\kappa},z)\}$.\cite{Men-13}
This case will be considered later on. Now we focus on the
so-called interface ordinary state with the envelope function
$\{\Phi_{o}(\bm{\kappa},z),\Theta_{o}(\bm{\kappa},z)\}$
corresponding to the natural boundary conditions:\cite{Men-13}
\begin{equation}\label{NBC-T}
i\frac{\delta
H_{t}(\bm{\kappa},-i\partial_{z})}{\delta(-i\partial_{z})}\Theta(\bm{\kappa},z)|_{z=0+}
-2dV(\bm{\kappa})\Phi(\bm{\kappa},z)|_{z=0-}=0,
\end{equation}
\begin{equation}\label{NBC-M}
i\frac{\delta
H_{m}(\bm{\kappa},-i\partial_{z})}{\delta(-i\partial_{z})}\Phi(\bm{\kappa},z)|_{z=0-}
-2dV^{\dag}(\bm{\kappa})\Theta(\bm{\kappa},z)|_{z=0+}=0.
\end{equation}
In the left half-space, the components of the envelope function
$\Phi_{o}(\bm{\kappa},z)$ are given by:
\begin{equation}\label{phi-o}
\phi^{(n)}(\bm{\kappa},z)=\phi^{(n)}(\bm{\kappa},0)\exp[p^{(n)}(\kappa)z],
\end{equation}
\begin{equation}\label{phi-o-0}
\phi^{(n)}(\bm{\kappa},0)=\frac{(-1)^{n+1}d}{\mathrm{N}p^{(n)}(\kappa)}
\sum_{j=1}^{4}V_{j}^{(n)\ast}(\kappa)\theta_{j}(\bm{\kappa},0),
\end{equation}
\begin{equation}\label{p1}
p^{(1)}(\kappa)=\sqrt{\kappa^{2}+[\Lambda+\mathrm{E}_{0}-E+\mathrm{M}]/\mathrm{N}},
\end{equation}
\begin{equation}\label{p2}
p^{(2)}(\kappa)=\sqrt{\kappa^{2}+[\Lambda-\mathrm{E}_{0}+E+\mathrm{M}]/\mathrm{N}},
\end{equation}
\begin{equation}\label{p3}
p^{(3)}(\kappa)=\sqrt{\kappa^{2}+[\Lambda+\mathrm{E}_{0}-E-\mathrm{M}]/\mathrm{N}},
\end{equation}
\begin{equation}\label{p4}
p^{(4)}(\kappa)=\sqrt{\kappa^{2}+[\Lambda-\mathrm{E}_{0}+E-\mathrm{M}]/\mathrm{N}},
\end{equation}
where $V_{j}^{(n)}(\kappa)$ are the matrix elements of the
hybridization potential $V(\mathbf{r})$, the subscript $j$ and
superscript $(n)$ are related to the TI and FMI states,
respectively.

Inserting $\Phi(\bm{\kappa},0)$ Eq.~(\ref{phi-o-0}) into
Eq.~(\ref{NBC-T}) one arrives at the relation for the interface
magnitude of the ordinary envelope function on the TI side:
\begin{equation}\label{NBC-TT}
i\frac{\delta
H_{t}(\bm{\kappa},-i\partial_{z})}{\delta(-i\partial_{z})}\Theta(\bm{\kappa},0)
-2dU(\bm{\kappa},E)\Theta(\bm{\kappa},0)=0,
\end{equation}
which contains the effective local pseudo-potential
$U(\bm{\kappa},E)$ seen by electrons on the TI side. The potential
matrix elements have the form
\begin{equation}\label{EffPP-U}
U_{jj'}(\bm{\kappa},E)=\sum_{n=1}^{4}(-1)^{n+1}U_{jj'}^{(n)}(\bm{\kappa},E),
\end{equation}
\begin{equation}\label{EffPP-U-jj}
U_{jj'}^{(n)}(\bm{\kappa},E)=\frac{dV_{j}^{(n)}V_{j'}^{(n)\ast}}{\mathrm{N}p^{(n)}(\kappa,E)}.
\end{equation}

The matrix elements $U_{jj'}^{(n)}(\bm{\kappa},E)$ characterize
internal properties of the interface. In the case when
$V^{(n)}_{1,3}\neq 0$, $V^{(n)}_{2,4}=0$, the effective
pseudo-potential $U(\bm{\kappa},E)$ has four non-zero components:
$U_{11}(\bm{\kappa},E)=P(\bm{\kappa},E)+Q(\bm{\kappa},E)$,
$U_{33}(\bm{\kappa},E)=P(\bm{\kappa},E)-Q(\bm{\kappa},E)$,
$U_{13}(\bm{\kappa},E)=U_{31}^{\ast}(\bm{\kappa},E)$, where
$P(\bm{\kappa},E)$, $Q(\bm{\kappa},E)$ and $U_{13}(\bm{\kappa},E)$
are the potential, exchange and spin orbit contribution(s),
respectively. These expressions are derived under the condition
$|U_{13}(\bm{\kappa},E)|<<|Q(\bm{\kappa},E)|<<|P(\bm{\kappa},E)|$;
note, that $Q(\bm{\kappa},E)$ is proportional to the intrinsic
exchange potential $\mathrm{M}$ of the FMI Hamiltonian,
Eq.~(\ref{H-MI}). In the following, for the sake of simplicity, we
omit the dependence of $P(\bm{\kappa},E)$, $Q(\bm{\kappa},E)$ and
$U_{13}(\bm{\kappa},E)$ on both $\bm{\kappa}$ and $E$. Then, if
the hybridization with the FMI conduction band is predominant, one
obtains the estimation
\begin{equation}\label{P-and-Q}
P\simeq\frac{d|V|^{2}}{\mathrm{N}\sqrt{\Lambda+\mathrm{E}_{0}}},~
Q\simeq\frac{-d|V|^{2}\mathrm{M}}{2\mathrm{N}(\sqrt{\Lambda+\mathrm{E}_{0}})^{3}},~
U_{13}\simeq\frac{dV_{1}^{(1)}V_{3}^{(1)\ast}}{\mathrm{N}\sqrt{\Lambda+\mathrm{E}_{0}}},
\end{equation}
where $\Lambda+\mathrm{E}_{0}>>\mathrm{M}$,
$|V^{(1)}_{1}|=|V^{(3)}_{3}|=|V|$. The influence of the
off-diagonal component of the hybridization, $U_{13}$, on the
interface electron states has been discussed in
Ref.~\onlinecite{Men-13}. This influence is insignificant as the
interface spin-flip processes generating the component $U_{13}$
are relatively weak, $|V_{3}^{(1)}|\approx |V_{1}^{(3)}|<<|V|$. We
are interested in the investigation of the interface state spin
polarization along the $z$ axis. Therefore in the following,
without the loss of generality, we assume $U_{13}=0$. As a
consequence, the matrix $\mathcal{K}^{\alpha\beta}$ acquires the
diagonal form and the $z$-polarized exchange field is
$\Delta^{z}(z)=\mathcal{K}^{zz} s^{z}(z)$. Below the upper indices
can be omitted, i.e., $\Delta(z)=\mathcal{K} s(z)$.

In general case, under the nontrivial boundary conditions and the
self-consistent potentials $\varphi(z)$ and $\Delta(z)$, we meet a
problem to seek for the solution of the non-linear equation
$[H_{t}(\bm{\kappa},-i\partial_{z})-E]\Theta(\bm{\kappa},z)=0$ (at
$z>0$). It is evident, no exact analytical solution for such the
task is available. Therefore, to capture the principal features of
the solution, we restrict ourselves to the lowest order of
perturbation theory in the interface potential and disregard the
feedback influence of these potentials on the envelope function
and spectrum the ordinary bound state. As it will be shown below,
the weak interface potential approximation, even being relatively
simplified, gives a good opportunity to clearly understand what
role the ordinary bound state plays in the establishing of an
exchange field on the TI side of the contact.

Following the procedure of Ref.~\onlinecite{Men-13} one can obtain
the expressions for the spectrum $E_{o}(\kappa)$ of the interface
ordinary state and the envelope function
$\Theta_{o}(\bm{\kappa},z)$ at $z>0$. If the interface potential
is weak, $|P,Q|/\Xi<<1$, the ordinary state spectrum for small
momenta is given by the transparent formula:
\begin{equation}\label{spectrum}
E_{o}^{(\pm)}(\kappa)=
-\widetilde{P}\pm\sqrt{\widetilde{Q}^{2}+\mathrm{A}^{2}\kappa^{2}},
\end{equation}
\begin{equation}\label{P-and-Q-til}
\widetilde{P}=\frac{2dP\sqrt{\Xi}}{\sqrt{\mathrm{B}}}\frac{\sqrt{\lambda}}{1+\lambda},~
\widetilde{Q}=\frac{2dQ\sqrt{\Xi}}{\sqrt{\mathrm{B}}}\frac{\sqrt{\lambda}}{1+\lambda},
\end{equation}
where $\lambda=\mathrm{A}^{2}/4\mathrm{B}\Xi$ is the parameter of
the TI bulk band spectrum, which is implied to be $\lambda
\geqslant 1$. The spin-independent part of the interface potential
gives the energy shift $-\widetilde{P}$ to the dispersion
relation, while the $z$ component of the exchange part of the
interface potential causes the energy gap of the size
$2\widetilde{Q}$ at the node point; the label $\pm$ in
Eq.~(\ref{spectrum}) distinguishes the states above and below the
gap.

Keeping the first order in the terms of the interface potential,
after some algebra, the envelope function of the ordinary state
with spin projection $\sigma=\pm$,
$\Theta_{o}^{\sigma}(\bm{\kappa},z)$, is reduced to rather simple
form
\begin{eqnarray}\label{EF-ordin}
\Theta_{o}^{\sigma}(\bm{\kappa},z)&=&
C_{o}^{\sigma}(\kappa)\{\Theta_{1}^{\sigma}(\phi)\exp[-q_{1}(\kappa)z]
\\&+&\Theta_{2}^{\sigma}(\phi)\exp[-q_{2}(\kappa)z]\}\nonumber,
\end{eqnarray}
\begin{eqnarray}\label{EF-ordin-1}
\Theta_{1}^{\sigma}(\phi)&=&(i(1+\alpha_{1}^{\sigma}),sign(\mathrm{A})(1+\alpha_{2}^{\sigma}),\\&
& e^{i\phi}(1+\alpha_{1}^{-\sigma}),
sign(\mathrm{A})ie^{i\phi}(1+\alpha_{2}^{-\sigma}))^{\mathrm{T}}\nonumber,
\end{eqnarray}
\begin{eqnarray}\label{EF-ordin-2}
\Theta_{2}^{\sigma}(\phi)&=&(i(1+\beta_{1}^{\sigma}),sign(\mathrm{A})(1+\beta_{2}^{\sigma}),\\&
& e^{i\phi}(1+\beta_{1}^{-\sigma}),
sign(\mathrm{A})ie^{i\phi}(1+\beta_{2}^{-\sigma}))^{\mathrm{T}}\nonumber,
\end{eqnarray}
where $C_{o}^{\sigma}(\kappa)$ is normalization factor. The
corrections $\alpha_{j}^{\sigma}$ and $\beta_{j}^{\sigma}$ satisfy
the relations:
\begin{eqnarray}\label{links}
-\widetilde{P}+\sigma
\widetilde{Q}&=&|\mathrm{A}|q_{1}(\alpha_{1}^{\sigma}-\alpha_{2}^{\sigma})
=|\mathrm{A}|q_{2}(\beta_{1}^{\sigma}-\beta_{2}^{\sigma})\\&=&\nonumber
|\mathrm{A}|[q_{1}-q_{2}](\beta_{2}^{\sigma}-\alpha_{2}^{\sigma})/4\\&=&\nonumber
\mathrm{B}\Xi[q_{1}-q_{2}](\beta_{1}^{\sigma}-\alpha_{1}^{\sigma})/|\mathrm{A}|
\nonumber.
\end{eqnarray}
The characteristic momenta are given by
\begin{equation}\label{q1}
q_{1}(\kappa)=\frac{|\mathrm{A}|+\sqrt{\mathrm{A}^{2}-4\mathrm{B}\Xi(\kappa)}}{2\mathrm{B}},
\end{equation}
\begin{equation}\label{q2}
q_{2}(\kappa)=\frac{|\mathrm{A}|-\sqrt{\mathrm{A}^{2}-4\mathrm{B}\Xi(\kappa)}}{2\mathrm{B}},
\end{equation}
and $q_{1,2}=q_{1,2}(0)$, $\Xi(\kappa)=\Xi-\mathrm{B}\kappa^{2}$.
We neglect a weak dependence of the pre-exponential factors
$\Theta_{1,2}^{\sigma}(\phi)$
(\ref{EF-ordin-1})-(\ref{EF-ordin-2}) on $\kappa$. The corrections
$\alpha_{j}^{\sigma}$ and $\beta_{j}^{\sigma}$ reflect the fact
that the interface potential lifts both the electron-hole
degeneration and the spin degeneration of the TI bulk Hamiltonian
(\ref{H-TI}). From Eq.~(\ref{links}) it is clear that, when $Q\neq
0$, one has $\alpha_{j}^{\sigma}\neq \alpha_{j}^{-\sigma}$ and
$\beta_{j}^{\sigma}\neq \beta_{j}^{-\sigma}$. Without the
interface potential the corrections are absent,
$\alpha_{j}^{\sigma}=0$ and $\beta_{j}^{\sigma}=0$. The
probability density of the ordinary state (\ref{EF-ordin}),
$|\Theta_{o}^{\sigma}(\bm{\kappa},z)|^{2}$, shows the maximum
value at $z=0$ and decays exponentially into TI with the
characteristic length $z_{o}\simeq q_{2}^{-1}$.

The charge density $n_{o}(z)$ and spin density $s_{o}(z)$
associated with the formation of the bound ordinary state on the
TI side of the TI/FMI interface may be determined as $n_{o}(z)=
\sum_{\bm{\kappa}}\Theta_{o}^{\dag}(\bm{\kappa},z)(\frac{\mathbb{I}}{2})\Theta_{o}(\bm{\kappa},z)$
and $s_{o}(z)=
\sum_{\bm{\kappa}}\Theta_{o}^{\dag}(\bm{\kappa},z)\frac{\tau_{0}\sigma_{z}}{2}\Theta_{o}(\bm{\kappa},z)$,
respectively, where the sum runs over occupied states. When the
dispersion law is expressed by
Eqs.~(\ref{spectrum})-(\ref{P-and-Q-til}), one can write these
densities through the squared components of the envelope function
(\ref{EF-ordin}) as
\begin{eqnarray}\label{charge}
n_{o}(z)&=&\frac{a^{2}}{4\pi
\mathrm{A}^{2}}\int_{-W}^{\mu}dE(E+\widetilde{P})
\\&\times&\nonumber
\biggl[h(E+\widetilde{P}-|\widetilde{Q}|)+h(-E-\widetilde{P}-|\widetilde{Q}|)\biggr]\\&\times&\nonumber
\sum_{j=1}^{4}|\theta_{oj}^{+}(\bm{\kappa},z)|^{2}\nonumber,
\end{eqnarray}
\begin{eqnarray}\label{spin-pol-01}
s_{o}(z)&=&\frac{a^{2}}{4\pi
\mathrm{A}^{2}}\int_{-W}^{\mu}dE(E+\widetilde{P})
\\&\times&\nonumber
\biggl[h(E+\widetilde{P}-|\widetilde{Q}|)+h(-E-\widetilde{P}-|\widetilde{Q}|)\biggr]\\&\times&\nonumber
\biggl[|\theta_{o1}^{+}(\bm{\kappa},z)|^{2}-|\theta_{o2}^{+}(\bm{\kappa},z)|^{2}\\&+&
\nonumber
|\theta_{o2}^{-}(\bm{\kappa},z)|^{2}-|\theta_{o1}^{-}(\bm{\kappa},z)|^{2}\biggr]\nonumber,
\end{eqnarray}
where $\mu$ is the chemical potential, $W$ is a cut-off energy,
$W\simeq|\mathrm{A}|a^{-1}\simeq \Xi$, $a$ is in-plane lattice
constant. In Eqs.~(\ref{charge}) and (\ref{spin-pol-01}), we have
used the above relations (\ref{links}).

The FMI magnetization opens the gap $2|\widetilde{Q}|$ in the
ordinary state spectrum at the node point $E_{o}(0)$ being
somewhat remote from the chemical potential $\mu$. Here we study
the relevant regime of $\mu+\widetilde{P}>|\widetilde{Q}|$. In the
leading non-vanishing order in the interface energies, which are
far below the bulk energy gap, $\widetilde{P}<<\Xi$ and
$|\widetilde{Q}|<<\Xi$, it is now straightforward to use
Eqs.~(\ref{charge}) and (\ref{spin-pol-01}) to find that
\begin{equation}\label{charge-dens}
n_{o}(z)\simeq \frac{2a^{2}|C_{o}|^{2}}{\pi
\mathrm{A}^{2}}(W^{2}+\mu^{2})g_{o}(z),
\end{equation}
\begin{equation}\label{g-ord}
g_{o}(z)=\frac{1}{4}\biggl\{\exp[-q_{1}z]+\exp[-q_{2}z]\biggr\}^{2},
\end{equation}
\begin{equation}\label{spin-dens-01}
s_{o}(z)\simeq \frac{2a^{2}|C_{o}|^{2}}{\pi
\mathrm{A}^{2}}\frac{\widetilde{Q}}{\Xi}(W^{2}+\mu^{2})f(z),
\end{equation}
\begin{eqnarray}\label{f}
f(z)&=&\frac{\Xi}{2|\mathrm{A}|}\biggl\{\exp[-q_{1}z]+\exp[-q_{2}z]\biggr\}\\&\times&\nonumber
\biggl\{\frac{\exp[-q_{1}z]}{q_{1}}+\frac{\exp[-q_{2}z]}{q_{2}}\biggr\}\nonumber,
\end{eqnarray}
where
$|C_{o}|^{2}=|C_{o}(0)|^{2}=|\mathrm{A}|/8\mathrm{B}(1+\lambda)$,
$g_{o}(0)=f(0)=1$. Since the dependence of the function
$\theta_{oj}^{\pm}(\bm{\kappa},z)$ on $\bm{\kappa}$ is weak
enough, it has been justified to set $\bm{\kappa}=0$ in the
integrand in Eqs.~(\ref{charge}) and (\ref{spin-pol-01}).

\begin{figure}
\includegraphics[width=\columnwidth]{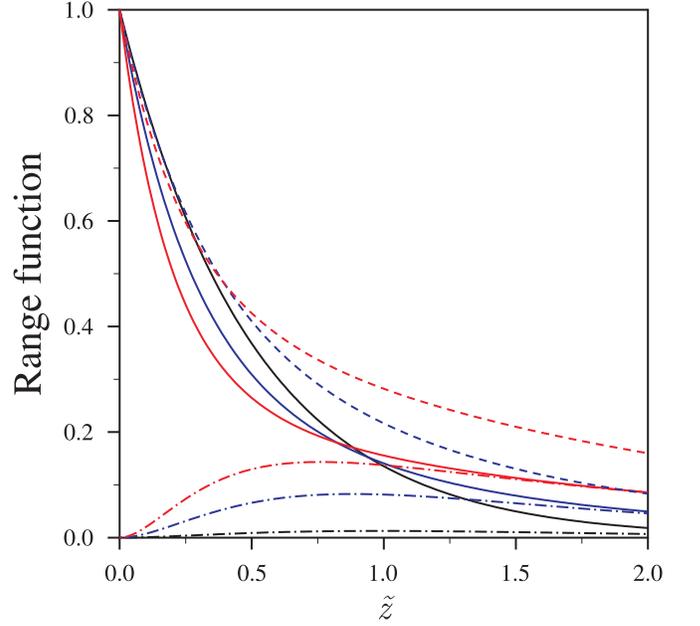}
\caption{(Color online) The range functions of the charge and spin
densities for the ordinary state, $g_{o}(z)$ (solid line) and
$f(z)$ (dashed line), at given band parameter $\lambda=1.0$ (black
line), $\lambda=2.0$ (blue lines), $\lambda=4.0$ (red lines). Note
that at $\lambda=1.0$ the dependences $g_{o}(z)$ and $f(z)$
coincide.  The range function of the topological state density,
$g_{t}(z)$ (dotted-dashed line), at given band parameter
$\lambda=1.1$ (black line), $\lambda=2.0$ (blue lines),
$\lambda=4.0$ (red lines).
$\widetilde{z}=z\sqrt{\frac{\Xi}{\mathrm{B}}}$ is the
dimensionless coordinate.}
 \label{fig}
\end{figure}

Thus, when TI is brought into contact with FMI, the hybridization
between the orbitals of TI and FMI at the interface induces the
spin polarized ordinary bound state on the topological side of the
contact. The direction of spin polarization of this state is
opposite to the FMI magnetization, the magnitude $s_{o}(0)$ is
proportional to $\widetilde{Q}\sim\mathrm{M}$ and depends on the
occupation of the state. Thank to the ordinary spin polarized
state, the induced exchange field $\Delta_{o}(z)=\mathcal{K}
s_{o}(z)$ penetrates inside  TI on the length scale far above the
lattice spacing, $z_{o}/2>>a,d$. The spin and charge  spatial
distributions, $g_{o}(z)$ Eq.~(\ref{g-ord}) and $f(z)$
Eq.~(\ref{f}), are illustrated in Fig.~\ref{fig}.

%%%%%%%%%%%%%%%%%%%%%%%%%%%%%%%%%%%%%%%%%%%%%%%%%%%%%%%%%%%%%%%%%%%%%%%%%%%%%%%%%%%%%%%%%%%%%%%%%%%%%%%%%%%%%
%%%%%%%%%%%%%%%%%%%%%%%%%%%%%%%%%%%%%%%%%%%%%%%%%%%%%%%%%%%%%%%%%%%%%%%%%%%%%%%%%%%%%%%%%%%%%%%%%%%%%%%%%%%%%

\section{TOPOLOGICAL BOUND STATE MODIFICATION}

As said above, the contact between TI and topologically trivial
insulator hosts, besides the ordinary bound state, the so-called
topological bound state, which is specified by the Dirac spectrum
and the envelope function
$\{\Phi_{t}(\bm{\kappa},z),\Theta_{t}(\bm{\kappa},z)\}$. In the
entire left half-space the envelope function is trivial,
$\Phi_{t}(\bm{\kappa},z)=0$, in the right half-space it conforms
to the conditions $\Theta_{t}(\bm{\kappa},z\rightarrow\infty)=0$
and $\Theta_{t}(\bm{\kappa},0+)=0$. Under these conditions the
solution of the equation
$[H_{t}^{0}(\bm{\kappa},-i\partial_{z})-E]\Theta(\bm{\kappa},z)=0$
(at $z>0$) has the form
\begin{eqnarray}\label{topo-func-0}
\Theta_{t}^{0(\pm)}(\bm{\kappa},z)&=&
C_{t}(\kappa)\Theta^{0(\pm)}(\phi)
\\&\times&\{\exp[-q_{1}(\kappa)z]-\exp[-q_{2}(\kappa)z]\}\nonumber,
\end{eqnarray}
where the spinor $\Theta^{0(\pm)}(\phi)=(i,sign(\mathrm{A}),\pm
e^{i\phi},\pm sign(\mathrm{A})ie^{i\phi})^{\mathrm{T}}$ depends
only on the polar angle of the momentum, $k_{\pm}=\kappa e^{\pm
i\phi}$. The functions $\Theta_{t}^{0(+)}(\bm{\kappa},z)$ and
$\Theta_{t}^{0(-)}(\bm{\kappa},z)$ describe the states with the
positive $E_{t}^{0(+)}(\kappa)=|\mathrm{A}|\kappa$ and negative
$E_{t}^{0(-)}(\kappa)=-|\mathrm{A}|\kappa$ energy, respectively;
$C_{t}(\kappa)$ is the normalization factor.

By analogy with Eq.~(\ref{charge-dens}) one can to write the
expression for the charge density in the unperturbed topological
state (\ref{topo-func-0}):
\begin{equation}\label{charge-dens-topo}
n_{t}(z)\simeq \frac{2a^{2}|C_{t}|^{2}}{\pi
\mathrm{A}^{2}}(W^{2}+\mu^{2})g_{t}(z),
\end{equation}
\begin{equation}\label{g-topo}
g_{t}(z)=
\frac{1}{4}\biggl\{\exp[-q_{1}z]-\exp[-q_{2}z]\biggr\}^{2},
\end{equation}
where $|C_{t}|^{2}=|\mathrm{A}|/8\mathrm{B}(\lambda-1)$. The
spatial dependence (\ref{g-topo}) is given in Fig. 1. The
topological state decays into the TI bulk on the length scale
$z_{o}/2$, but the maximum of the density $n_{t}(z)$ does not
occur at the interface, where $n_{t}(0)=0$, but rather near the
point $z_{t}=\ln(q_{1}/q_{2})/(q_{1}-q_{2})$ ($z_{t}\lesssim
\sqrt{\frac{\mathrm{B}}{\Xi}}<z_{o}$) on the TI side that is
distant from the interface. Therefore, within our continual model,
the topological state is directly insensitive to the local
effective interface potential. Correspondingly, the direct
magnetic coupling between the topological state and the FMI
magnetization is absent since this state is spatially separated
from the interface. Nevertheless, the topological state is
subjected to the indirect influence of the interface with FMI
through the extended fields $\varphi_{o}(z)$ and $\Delta_{o}(z)$
induced inside the TI host due to the orbital intermixing at the
interface. We further show how the embedded exchange field
$\Delta_{o}(z)$ affects the energy spectrum of the topological
state.

Strictly speaking, we ought to find out an evanescent solution of
the equation
$[H_{t}(\bm{\kappa},-i\partial_{z})-E]\Theta(\bm{\kappa},z)=0$ at
$z\rightarrow\infty$ and $z=0$, wherein the Hamiltonian operator
(\ref{H-T-til}) has the potential energy
$\mathbb{U}=\varphi(z)\mathbb{I}+
\tau_{0}\otimes(\bm{\sigma}\cdot\bm{\Delta}(z))$ with rather
complicated matrix and spatial dependences. In the context of the
magnetic proximity effect, we are interested to know how the
induced exchange field $\Delta_{o}(z)$, applied along the $z$
axis, affects the electron spectrum of the topological state.
Therefore we treat the potential energy $\mathbb{U}$ as a
perturbation and remain only the exchange part
$\mathbb{U}_{ex}(z)=\tau_{0}\otimes\sigma_{z}\Delta_{o}(z)$, where
$\Delta_{o}(z)=\mathcal{K} s_{o}(z)$, the function $s_{o}(z)$ is
given by Eqs.~(\ref{spin-dens-01}) and (\ref{f}).

To estimate the modification of the topological state near the
Dirac point under the exchange field, we utilize the method
similar to the perturbation theory treatment for electron terms
with close eigenenergies.\cite{Landau} Indeed, near the Dirac
point $\kappa=0$, there are energies in the spectrum of the
unperturbed topological state,
$E_{t}^{0(\pm)}(\kappa)=\pm|\mathrm{A}|\kappa$, the difference
between which does not exceed the perturbation value
$|\Delta_{o}(z)|$. For the perturbed envelope function
$\Theta(\bm{\kappa},z)$ we employ the ansatz having the same
spatial dependence of Eq.~(\ref{topo-func-0}), but the variable
spinor structure
$\Theta^{(\pm)}=(\theta_{1},\theta_{2},\theta_{3},\theta_{4})^{\mathrm{T}}$
that is driven by the perturbation. To seek for the factors
$\theta_{j}$ we calculate the integral
\begin{equation}\label{integral-0}
\int_{0}^{\infty}dz\Theta^{\dag}(\bm{\kappa},z)
[H_{t}^{0}(\bm{\kappa},-i\partial_{z})+\mathbb{U}_{ex}(z)-E]
\Theta(\bm{\kappa},z)=0.
\end{equation}
The corresponding secular equation yields the spectrum of the
topological state under the embedded exchange field:
\begin{equation}\label{spectrum-topo}
E_{t}^{(\pm)}(\kappa)=\pm\sqrt{\mathrm{A}^{2}\kappa^{2}+I^{2}\Delta_{o}^{2}(0)},
\end{equation}
where
\begin{equation}\label{overlap}
I=\int_{0}^{\infty}dzf(z)[\Theta_{t}^{0(\pm)}(0,z)]^{\dag}\mathbb{I}\Theta_{t}^{0(\pm)}(0,z)
=\frac{1}{2}\frac{3\lambda-1}{3\lambda+1}.
\end{equation}
The overlap integral $I$ is a function only of the material
parameters of the TI bulk. The perturbation violates a parity
between the minority and majority spin orientations of the
electron states along the $z$ axis in accordance with
\begin{eqnarray}\label{spinor-topo-0}
\Theta^{(\pm)}(\bm{\kappa})&\simeq& \biggl(i,sign(\mathrm{A}),\\&&
\frac{e^{i\phi}|\mathrm{A}|\kappa}{E_{t}^{(\pm)}(\kappa)-I\Delta_{o}(0)},
\frac{ie^{i\phi}\mathrm{A}\kappa}{E_{t}^{(\pm)}(\kappa)+I\Delta_{o}(0)}\biggr)^{\mathrm{T}}\nonumber.
\end{eqnarray}

Ergo, the induced exchange field associated with the ordinary
state penetrates into the TI host over distance on the order of
$z_{o}/2$ to break the time reversal symmetry. As a consequence,
the energy gap opens in the Dirac spectrum of the interface
topological state $E_{t}^{(\pm)}(\kappa)$ (\ref{spectrum-topo}).
The gap size is directly proportional to the FMI magnetization and
determined by the overlap of the ordinary state spin polarization
and the topological state electron density, Eq.~(\ref{overlap}).
As follows from Eqs.~(\ref{P-and-Q}), (\ref{spin-dens-01}) and
(\ref{overlap}), the induced gap size $2I\Delta_{o}(0)$ is limited
by a number of factors: the intermixing intensity of the TI and
FMI states at the interface, $Q$; the TI bulk band structure,
$\lambda$; the Fermi level position, $\mu$, defining the TI states
filling; and the exchange interaction strength in the TI bulk,
$\mathcal{K}$.

The fact used in the present work is that the topological and
ordinary states respond highly distinctly to the perturbation
created by the TI/FMI interface. We show that the gap opens at the
Dirac point of the topological state due to the exchange field
originated from the spin polarized ordinary state. We conclude
that the exchange coupling transfer through the mediation of the
ordinary state is a key aspect of the mechanism of achieving
interplay between FMI and the helical state in TI.

\section{SUMMARY AND CONCLUDING REMARKS}

In this study, we have succeeded in understanding of the physical
mechanism for magnetic proximity effect in the TI/FMI
heterostructures, by using a rather simple model for both
insulators and phenomenologically regarding the interface mixing
between their states but ignoring the fine details of the
interface on atomic scale. We have applied the envelope function
method to study the in-gap bound states at the TI/FMI interface,
wherein the narrow-gap semiconductor with inverted band structure
is in the contact with the wide-gap magnetic semiconductor with
normal band structure. Within the continual approach, to
analytically describe the ordinary and topological interface
states and interplay between them we have used the perturbation
theory in the terms of the interface potential and the local
approximation for an electron-electron interaction. It is no
reason to think that there will be a qualitative difference with
the obtained results when the self-consistent electrostatic and
exchange fields are taken into account in the presence of an
arbitrarily strong interface potential. Nevertheless the question
about the electron density rearrangement within the interface
region on the TI side is highly important to discuss it. It is
particularly specific of the prototypical 3D TIs belonging to the
Bi$_{2}$Se$_{3}$ family, which are narrow-gap semiconductors, that
the screening scale is typically on the order of a quintuple
thickness, i.e. $\approx$1 nm. It means that the fields
$\varphi(\mathbf{r})$ and $\bm{\Delta}(\mathbf{r})$ figuring in
Eq.~(\ref{H-T-til}), in strict sense, are neither long-range nor
short-range fields. Therefore each of the limit treatments, the
renormalization of the interface potential (that would be under
the condition $D\ll z_{o}/2$) or the semiclassical treatment (that
would be suited under the condition $D\gg z_{o}/2$), is the
approximation for the fields $\varphi(\mathbf{r})$ and
$\bm{\Delta}(\mathbf{r})$ that can yield only qualitative
estimation for the spatial variation of electron density near the
interface. To improve the description of the TI/FMI contact on the
specific scale $D\approx z_{o}/2$, it should be reasonable one to
utilize a fitting procedure with the sectionally continuous
functions $\varphi(z)$ and $\Delta(z)$, for example, in the form
of a rectangular or triangular potential well attached to the
interface, where the value $\varphi(0)$ may, in principle, exceed
the bulk energy gap and thus may significantly shift the energy
spectrum $E_{o}^{\sigma}(\kappa)$ of the ordinary states relative
to the bulk energy spectrum of TI. It is evident that such the
modification of the model cannot change the principle conclusion
about the mechanism of the magnetic proximity effect in the system
under investigation.

An exchange field inside TI may also be induced by placing TI in
contact with AFMI due to the presence of uncompensated
magnetization in the outermost AFMI layer. The recent density
functional theory (DFT) first-principles calculations for the
Bi$_2$Se$_3$/MnSe superlattice expounded in
Ref.~\onlinecite{Eremeev} has examined in detail the magnetic
proximity effect near the 3D TI/AFMI (MnSe) interface. It was
shown that the charge redistribution and mixing of the
Bi$_2$Se$_3$ and MnSe orbitals at the interface brings on drastic
modification of the electron structure with respect to the
pristine Bi$_2$Se$_3$ surface. The calculation results reveal the
presence of the ordinary state with probability maximum near the
interface plane. This state penetrates into the first interfacial
quintuple layer of TI. At the $\Gamma$ point, it appears in the
local bulk valence band gap owing to the near-interface band
bending of $\approx-0.8$ eV. This state is gapped (56 meV) and
spin polarized due to the hybridization of the TI and AFMI states.
On the other hand, the probability maximum of the Dirac
topological state relocates from the first quintuple layer to the
second one, leaving directly unattainable for a magnetic
perturbation from MnSe. The interface ordinary state mediates an
exchange coupling between AFMI and the topological state due to an
overlap of the topological and trivial interface states within the
first interfacial quintuple layer. The topological state acquires
the energy gap of $\approx$8.5 meV proportional to the overlap.

The magnetic proximity effect in the TI/magnetic insulator (FMI or
AFMI) hybrid structures is rather intricate phenomenon. The
analytical continual model developed here and the DFT results of
Ref.~\onlinecite{Eremeev} are in good qualitative agreement with
each other. They unveil the unique route for the penetration of
the exchange field into TI including three stages: the magnetic
insulator magnetization $\rightarrow$ the interface ordinary state
$\rightarrow$ the interface topological state.

It is likely that our results would be highly helpful for the
analysis of the feasibility of recently proposed unusual physical
effects in TIs and the ``tailor-made" structures on their base,
such as anomalous quantum Hall effect,\cite{Yu} magnetic monopole
imaging,\cite{Qi-09} topological contribution to the Faraday and
Kerr effects,\cite{Qi-08} inverse spin-galvanic
effect.\cite{Garate} Our findings could provide guidelines to
engineer spintronic device applications, for instance, the
TI-based \emph{p-n} junctions\cite{Wang} and memory device based
on the TI surface coated with a magnetic insulator
film.\cite{Fujita}

In summary, our analysis provides insight into the microscopic
mechanism of the proximity effect in the TI/FMI hybrid structure.
The nature of the proximity effect is tangled enough. The delicate
moment is the presence of the ordinary state as a mediator for the
spin polarization transmission over the interface from FMI to the
topological state. We have distinguished way for modifying the
spectrum of the topological state through the interface-induced
exchange field that breaks time reversal symmetry, giving rise to
the gap opening at the Dirac point in the topological state
spectrum.

\begin{acknowledgments}
We acknowledge partial support from the Basque Country Government,
Departamento de Educaci\'{o}n, Universidades e Investigaci\'{o}n
(Grant No. IT-366-07), the Spanish Ministerio de Ciencia e
Innovaci\'{o}n (Grant No. FIS2010-19609-C02-00), the Ministry of Education and
Science of Russian Federation (No. 2.8575.2013), and Russian
Foundation for Basic Researches (grant 13-02-00016).

\end{acknowledgments}

\end{document}